\documentclass[runningheads]{llncs}

\usepackage[T1]{fontenc}
\usepackage{graphicx}
\usepackage{amsmath}
\usepackage{booktabs}
\usepackage{url}

\begin{document}

\title{Evaluating Priority Dispatch for Premium Membership in On-demand Taxi-sharing}
\titlerunning{Temporal Trade-offs of Premium Membership}

\author{Nozomi Yamamoto\orcidID{0009-0005-5770-5524} \and Hiroki Sakaji\orcidID{0000-0001-5030-625X}  \\ \and Itsuki Noda\orcidID{0000-0003-1987-5336}}
\authorrunning{N.Yamamoto et al.}

\institute{Hokkaido University, Sapporo 060-0808, Japan \\
\email{yamamoto.nozomi.i2@elms.hokudai.ac.jp;\\ \{sakaji, i.noda\}@ist.hokudai.ac.jp}}

\maketitle

\begin{abstract}
In this paper, we analyze the dispatch-level feasibility of introducing premium membership into an on-demand taxi-sharing service using a sequential optimal insertion method. We propose two dispatch-level methods: deadline shortening for premium users, and deadline shortening with dedicated premium vehicles. We compare them with a baseline dispatch method through SUMO-based simulations under three demand scenarios: steady uniform demand, time-varying demand with a peak period, and urban-focused demand with concentrated peak traffic. 
Our results show that the suitable priority mechanism depends on operating conditions and premium ratio, suggesting that condition-dependent method switching may provide premium benefits while keeping non-premium-user burden smaller for overall optimization.

\end{abstract}

\section{Introduction}

On-demand taxi-sharing services support shared trips among users and aim to provide lower fares than conventional taxis. Such services combine the flexibility of taxis with the efficiency of shared public transport. In recent years, transit deserts, which are gaps between public transport supply and travel demand, have been discussed internationally. 
Such services are expected to help fill such gaps, though their effectiveness and equity implications depend on local conditions and institutional design~\cite{Jiao2013TransitDeserts,Chen2022TransitDesertsEurope}. They also often depend on subsidies or incentives to secure supply, making long-term profitability difficult to achieve~\cite{ITF2019AppBasedMobility,Fang2019PricesSubsidiesSharing}.
One possible design option is to provide a premium service to time-sensitive users at a higher fare. Unlike air or rail travel where premium services focus on comfort, a premium service in a taxi-sharing system can potentially change travel time itself by changing dispatch priority. It remains unclear whether such differences in travel time can be achieved in practice, and whether this can be accomplished without placing an excessive time burden on non-premium users.


In this paper, we examine the introduction of premium membership into on-demand taxi-sharing services. Specifically, we focus on whether a measurable time advantage can be created operationally, leaving fare design for open issue. Premium users are given priority so that their total travel time from demand generation to arrival can be shortened. We also examine the impact on non-premium users to identify conditions under which premium service substantially degrades their service quality. To examine this trade-off, we propose and compare two dispatch mechanisms by using SUMO-based simulations: (1) a method that introduces deadline shortening into the sequential optimal insertion method, and (2) a method that extends the first method by reserving part of the fleet for premium users.
The main contributions of this paper are as follows.
\vspace{-1mm}
\begin{itemize}
    \item We show that deadline shortening creates a travel-time gap mainly when the fleet is under pressure, while its effect becomes limited when vehicle capacity is sufficient.
    \item We show that dedicated premium vehicles can create a clearer gap under off-peak conditions, but at the cost of reducing the effective capacity available to non-premium users under congestion.
    \item We show that premium ratio and demand pattern affect both the choice of priority mechanism and the appropriate premium ratio, helping operators design dispatch rules for their operating conditions.
\end{itemize}


\section{Related Work}

Research on ride-hailing and ride-pooling has mainly developed around dynamic vehicle assignment. Santi et al. quantified pooling potential using shareability networks~\cite{Santi2014}, and Alonso-Mora et al. proposed a real-time optimization framework for large-scale ride-pooling~\cite{AlonsoMora2017}.
However, these studies generally assume homogeneous users and do not explicitly address service differentiation.

Although few studies examine a membership-based premium scheme of the type considered in this study,  the underlying issue is closely related to the price--quality-of-service (QoS) tradeoff in on-demand mobility. Prior studies have shown that ride-pooling adoption and platform performance are shaped by the tradeoff among price, travel time, service quality, and system efficiency~\cite{HansenSener2023,deRuijter2021,WangZhang2022}.
More explicit differentiation has also been studied through waiting-time differentiation, service quality contracts, and heterogeneous service classes~\cite{DeMunck2023,Beirigo2022,Ma2024}. These studies provide important foundations for differentiated on-demand mobility, but they mainly focus on pricing or theoretical service design and often abstract away from real-time spatial dispatch interactions.

In contrast, this study isolates the dispatch-level mechanism: before optimizing fares, we evaluate whether priority dispatch can produce a measurable travel-time advantage and what burden this imposes on non-premium users under different demand patterns~\cite{SozaParra2024}.
\section{Problem Formulation}

This study investigates the trade-off inherent in introducing premium membership to on-demand taxi-sharing services, using the Smart Access Vehicle Service (SAVS) — a real-time dispatch platform — as a case study.

\subsection{Premium and Non-Premium Users}
This study classifies users into premium and non-premium groups. Premium users are time-sensitive travelers willing to accept an additional burden, such as a higher fare, in exchange for shorter travel time. The purpose of this classification is to examine dispatch behavior under non-uniform user treatment.
Here, the value of premium membership is defined as a reduction in total travel time from demand generation to arrival. Therefore, the analysis considers both the benefit to premium users and the degradation experienced by non-premium users. A desirable premium system should provide clear premium-user benefits while keeping non-premium user burden small.

\subsection{Allocation Method}

\begin{figure}[htbp]
\centering
\includegraphics[width=0.75\textwidth]{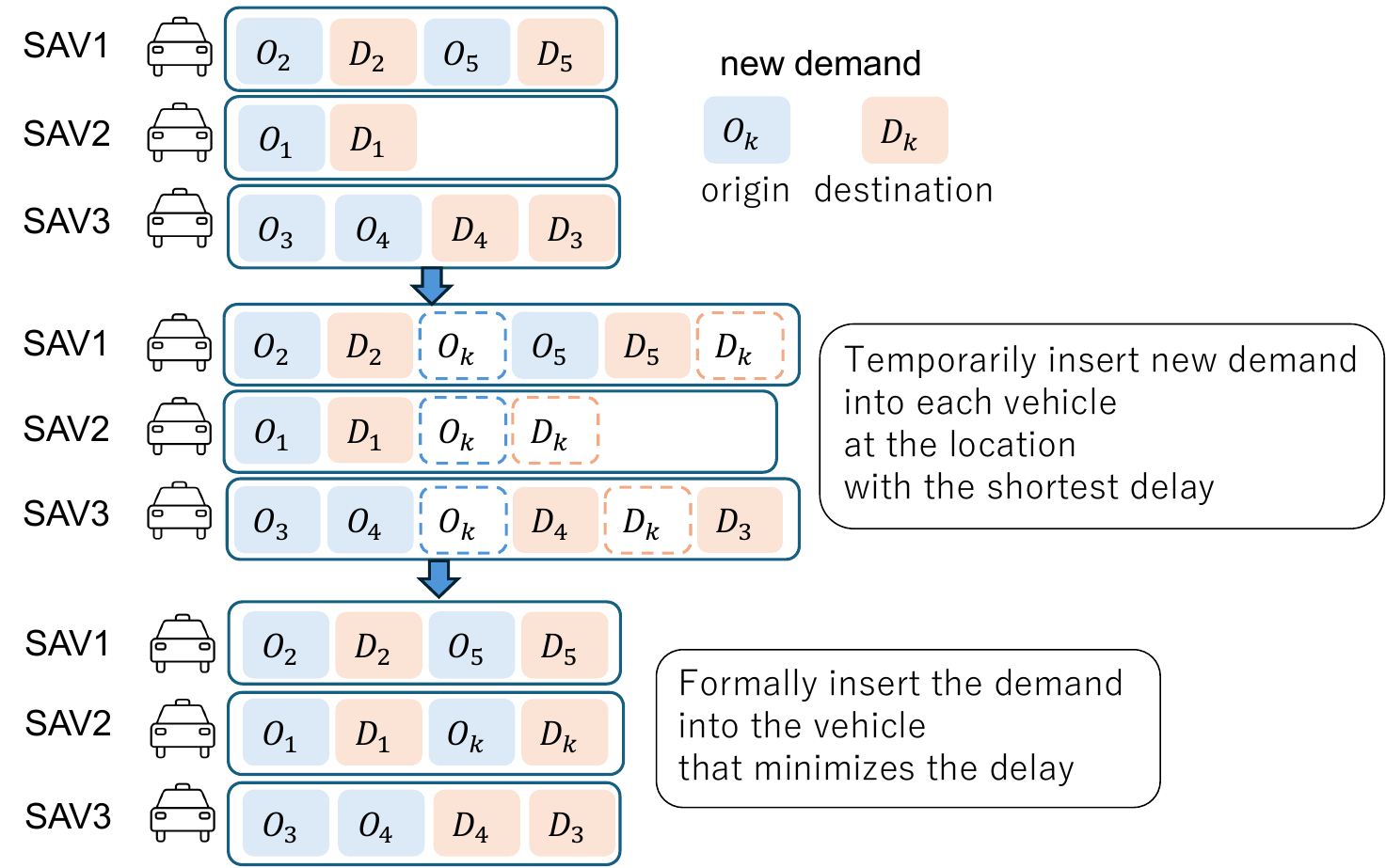}
\caption{Concept of the sequential optimal insertion method.}
\label{fig:insertion}
\end{figure}

SAVS uses the sequential optimal insertion method for dispatching. The method keeps, for each vehicle, an ordered list of waypoints corresponding to the origins and destinations of assigned demands. Once a waypoint sequence is stored, the order of existing waypoints is not changed. Fig.~\ref{fig:insertion} illustrates the concept.

The procedure is based on the method described by Noda et al.~\cite{noda2003evaluation}. 
First, each demand is divided into two waypoints: an origin and a destination. Each vehicle in the fleet maintains a waypoint list for the demands assigned to it. Second, each vehicle calculates the expected completion times of its assigned demands. Third, when a new demand appears, each vehicle tries to insert the origin and destination of the new demand into all feasible positions in its waypoint list. For each possible insertion, the system calculates the total delay caused at all waypoints from the expected completion times. The insertion cost is defined as this total delay plus the expected completion time of the newly inserted demand. Insertions that cause an existing or new demand to miss its deadline, defined in Section~4.1, are removed from consideration. Finally, the system compares the minimum insertion cost over the fleet and assigns the demand to the vehicle with the smallest cost.

\subsection{Efficiency Metric}

Travel time inherently depends on trip distance. Therefore, we use time per distance (TPD) as the primary metric, defined for each user $i$ as $\tau_i = (w_i + r_i) / d_i$, where $w_i$, $r_i$, and $d_i$ represent waiting time, in-vehicle travel time, and travel distance, respectively. Since waiting and travel times are only defined for completed trips, this metric is calculated exclusively for users whose demands were successfully fulfilled. We also use the cancel rate, defined as the proportion of canceled demands among all generated demands. In our analysis, we evaluate the performance using the ensemble average of time per distance, denoted as $\bar{\tau}^{P}$ for premium users and $\bar{\tau}^{R}$ for non-premium users. Specifically, we first calculate the mean $\tau_i$ for completed demands within each of the 30 independent simulation runs (varied by different random seeds), and then average these values across all runs to obtain the final representative metrics. A desirable outcome is one that minimizes $\bar{\tau}^{P}$ while preventing an excessive increase in $\bar{\tau}^{R}$. 


\section{Proposed Dispatch Methods}
This section explains the baseline and the two proposed methods. The baseline uses the ordinary sequential optimal insertion method without considering premium membership. Premium labels are retained only for evaluation, and both user groups use the same deadline setting.

\begin{figure}[tbp]
\centering
\includegraphics[width=0.8\textwidth]{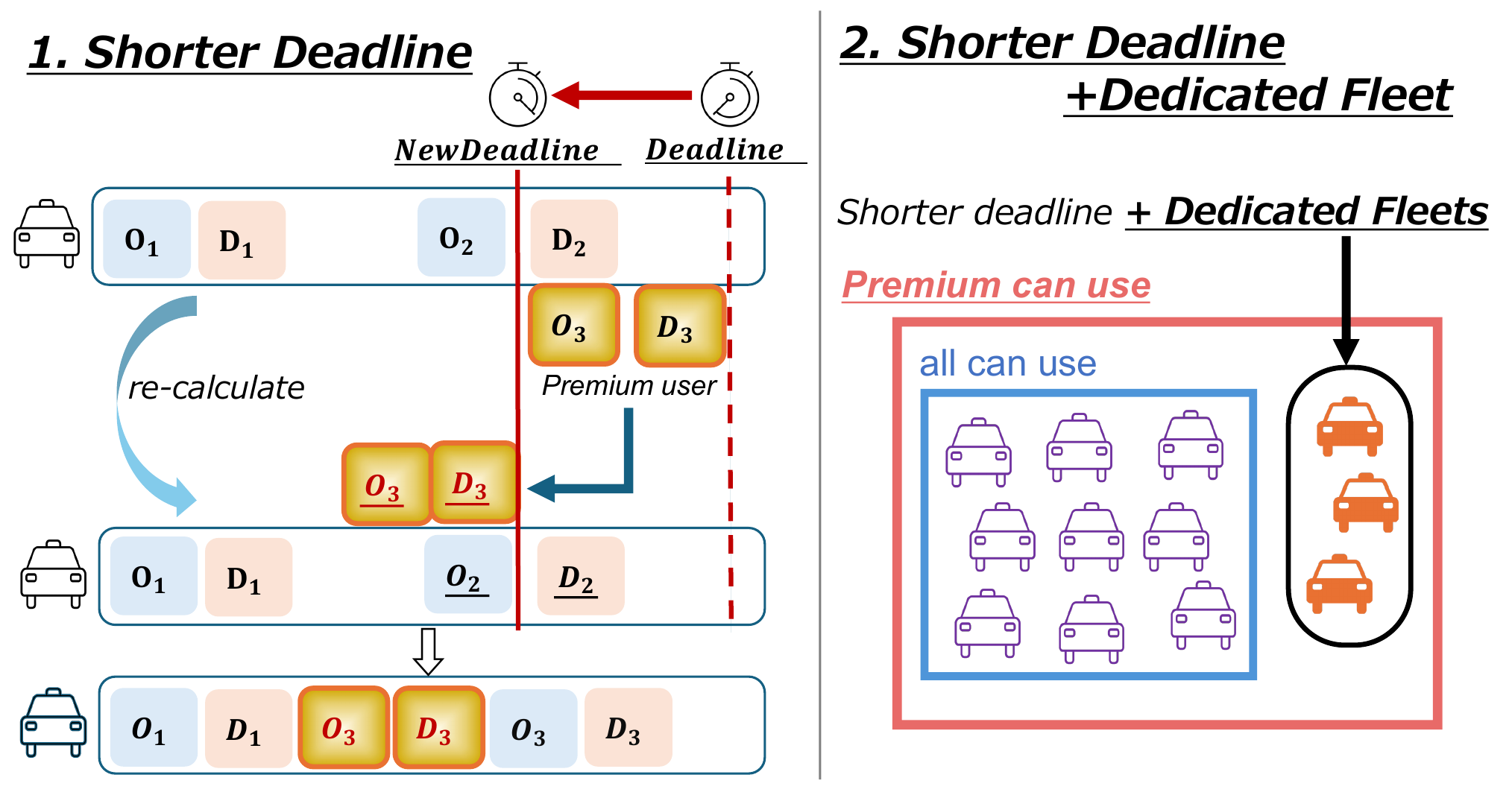}
\caption{Concept of the proposed dispatch methods. \textit{Shorter-Deadline} recalculates the allowable deadline interval for premium requests before insertion.}
\label{fig:proposed}
\end{figure}

\subsection{\textit{Shorter-Deadline}}


The sequential optimal insertion method incorporates a deadline variable that specifies the maximum allowable travel time from demand generation to arrival. Demands whose predicted arrival times exceed this deadline are rejected and treated as cancellations. This deadline is used as a dispatch feasibility constraint based on predicted travel time, not as an absolute guarantee of arrival time under all real-world traffic conditions.

In this study, we set the baseline allowable time interval $t_{\mathrm{deadline}}$ in proportion to trip distance, defined as:
$t_{\mathrm{deadline}} = \frac{d}{v_{\mathrm{slowest}}}$.
Here, $d$ is the estimated travel distance and $v_{\mathrm{slowest}}$ is a deadline slack parameter expressed in speed units.
The \textit{Shorter-Deadline} prioritizes premium users by applying a tighter time constraint. This is achieved by recalculating the deadline using an increased speed parameter $v_{\mathrm{prem}}$ ($> v_{\mathrm{slowest}}$). To avoid excessively strict deadlines, we incorporate the average vehicle speed $v_{\mathrm{sav}}$ and a lower-bound adjustment $N$, defining the premium deadline as: $t_{\mathrm{deadline}} = \max \left( \frac{d}{v_{\mathrm{prem}}}, \frac{d}{v_{\mathrm{sav}}} + N \right)$.
Here, $v_{\mathrm{slowest}}$ and $v_{\mathrm{prem}}$ function as slack parameters rather than physical speeds. Increasing $v_{\mathrm{prem}}$ strengthens priority by treating demands with higher urgency, though excessively strict deadlines may increase cancellations.
\subsection{Dedicated-Fleet}

\textit{Dedicated-Fleet} extends \textit{Shorter-Deadline} by dedicating part of the fleet to premium users. In this method, two types of vehicles are prepared: dedicated vehicles that can be used only by premium users, and shared vehicles that can be used by both premium and non-premium users. Premium demands are dispatched using both vehicle groups, whereas non-premium demands are dispatched using only shared vehicles. The deadline setting for premium users is the same as in \textit{Shorter-Deadline}.
Dedicated vehicles can assign premium users to less crowded vehicles, shortening their travel time and creating a clearer difference from non-premium users. At the same time, non-premium users can use fewer vehicles than in the baseline or \textit{Shorter-Deadline}.


\section{Experimental Settings}
This section presents the experimental settings. We first describe the map, fleet, and simulation horizon, then explain the demand scenarios, and finally define the evaluation metrics.
\subsection{Map, Fleet, and Simulation Horizon}
\begin{figure}[htbp]
\centering
\includegraphics[width=0.25\textwidth]{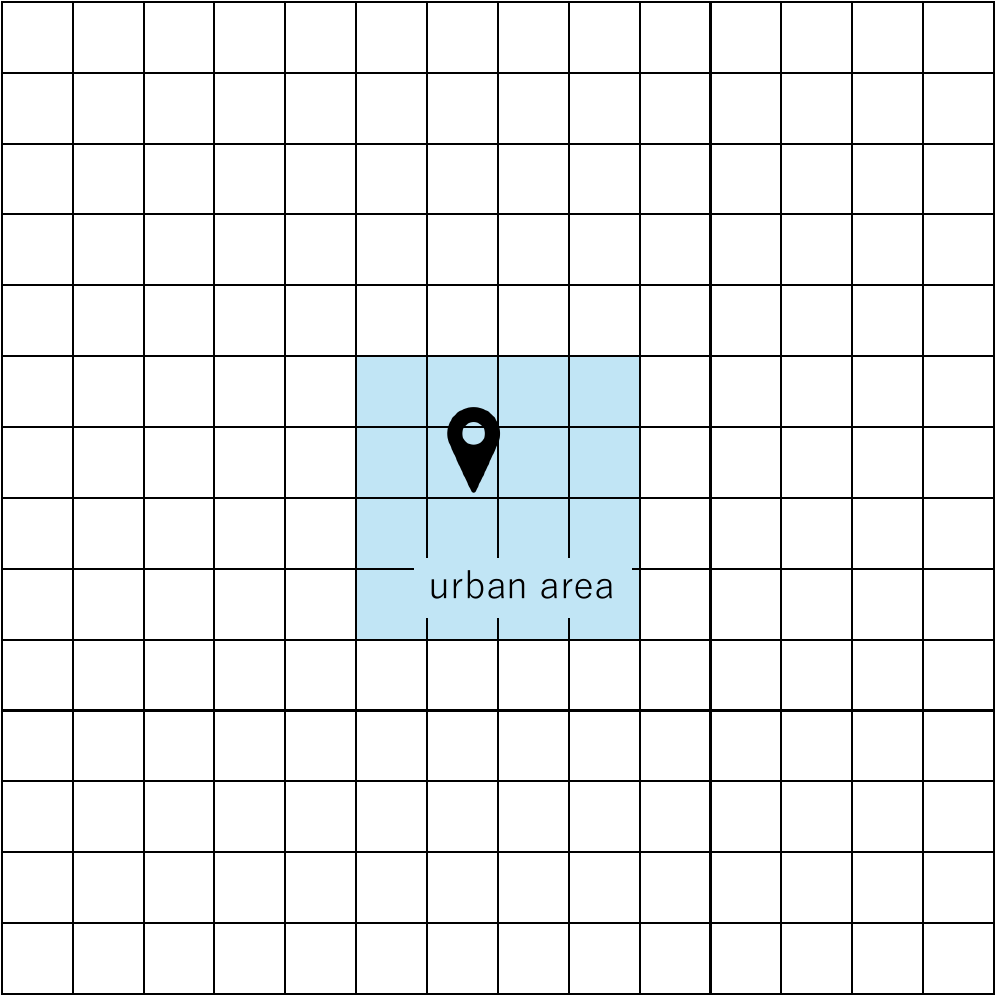}
\caption{Synthetic grid map used in the experiments.}
\label{fig:map}
\end{figure}

The experiments use a synthetic $15 \times 15$ grid map shown in Fig.~\ref{fig:map}. The node interval is 400~m, and the entire area is $5.6~\mathrm{km} \times 5.6~\mathrm{km}$. An urban area of $1.6~\mathrm{km} \times 1.6~\mathrm{km}$ is defined at the center of the map. The simulator uses two zones: entire area, which covers the whole map, and urban area, which covers the central area.

We compare three methods: Baseline, \textit{Shorter-Deadline}, and \textit{Dedicated-Fleet}.
In Baseline and \textit{Shorter-Deadline}, all 12 vehicles are shared by all users. In \textit{Dedicated-Fleet}, $\lfloor ratio \times savN \rfloor$ of the 12 vehicles are dedicated to premium users.
Premium users are assigned randomly at a fixed ratio among all demands. 
Each condition is simulated with 30 random seeds, and methods are compared under the same condition and the same seed. For details, see Table \ref{tab:fleet}.

\begin{table}[hhtbp]
\small
\centering
\caption{Fleet and parameter settings. All deadline terms are converted to seconds in the implementation.}
\label{tab:fleet}
\begin{tabular}{ll}
\toprule
Parameter & Value \\
\midrule
Number of vehicles $savN$ & 12 \\
Vehicle deployment interval & 60~s \\
Average vehicle speed $v_{\mathrm{sav}}$ & 30~km/h \\
Baseline deadline slack parameter $v_{\mathrm{slowest}}$ & 0.1~m/s \\
Premium deadline slack parameter $v_{\mathrm{prem}}$ & 0.1, 0.3, 0.5, 0.7, 1.0~m/s \\
Lower-bound adjustment value $N$ & 600~s \\
Premium ratio $ratio$ & 10\%,20\%,30\%,40\%,50\% \\
Simulation time & 43,200~s \\
Demand generation period & 1,800--37,800~s \\
Peak period & 16,200--23,400~s \\
Random seeds & 30 \\
\bottomrule
\end{tabular}
\end{table}




\subsection{Demand Scenarios}

This study evaluates three demand scenarios. S1 (\textit{steady}) has uniform demand over the entire area. S2 (\textit{time-varying}) has uniform demand over the entire area and a higher demand rate during the peak period. S3 (\textit{urban-focused}) has a center-focused demand structure and adds peak demand related to the urban area. Table~\ref{tab:demand} summarizes the scenarios.


\begin{table}[htbp]
\centering
\caption{Summary of Demand Scenarios (S1--S3). OD denotes an origin--destination zone pair. Approximate rates result from combining OD-specific demand streams.}
\label{tab:demand}

\fontsize{7pt}{8pt}\selectfont

\begin{tabular}{@{}lllc p{3.5cm}@{}}
\toprule
\textbf{Scenario} & \textbf{Phase} & \textbf{Distribution} &
\textbf{Req./h} & \textbf{OD Composition} \\ 
\midrule
\textbf{S1} & \textit{off-peak} & Uniform & 150 & All $\to$ All \\
\midrule
\textbf{S2} & \textit{off-peak} & Uniform & 150 & All $\to$ All \\
            & \textit{peak}   & Uniform & 300 & All $\to$ All \\
\midrule
\textbf{S3} & \textit{off-peak} & Center-focused & $\sim$151 &
Center $\to$ Center (30\%)\newline
Periph. $\to$ Center (30\%)\newline
Center $\to$ Periph. (20\%)\newline
Periph. $\to$ Periph. (20\%) \\
\cmidrule{2-5}
            & \textit{peak} & Center-heavy & $\sim$300 &
Center $\to$ Center (40\%)\newline
Periph. $\to$ Center (30\%)\newline
Center $\to$ Periph. (20\%)\newline
Periph. $\to$ Periph. (10\%) \\
\bottomrule
\end{tabular}

\end{table}

\subsection{Evaluation Metrics}
The main metric are TPD and cancel rate, as defined in Section~3.3. We evaluate premium users' improvement and non-premium users' degradation from the baseline, where positive values indicate a reduction in premium users' TPD and an increase in non-premium users' TPD. Additionally, we report the service gap $\delta$, defined as $\delta = \bar{\tau}^{R} - \bar{\tau}^{P}$.




\section{Results}
This section presents the simulation results. We first examine $\delta$. We then compare the average TPD for each scenario and method at $v_{\mathrm{prem}}=0.5$, examine cancellations, and Finally, we compare premium user benefit and non-premium user degradation in the trade-off space.

\subsection{Effect of \textit{Shorter-Deadline} and \textit{Dedicated-Fleet}}

\begin{figure}[htbp]
\centering
\includegraphics[width=0.75\textwidth]{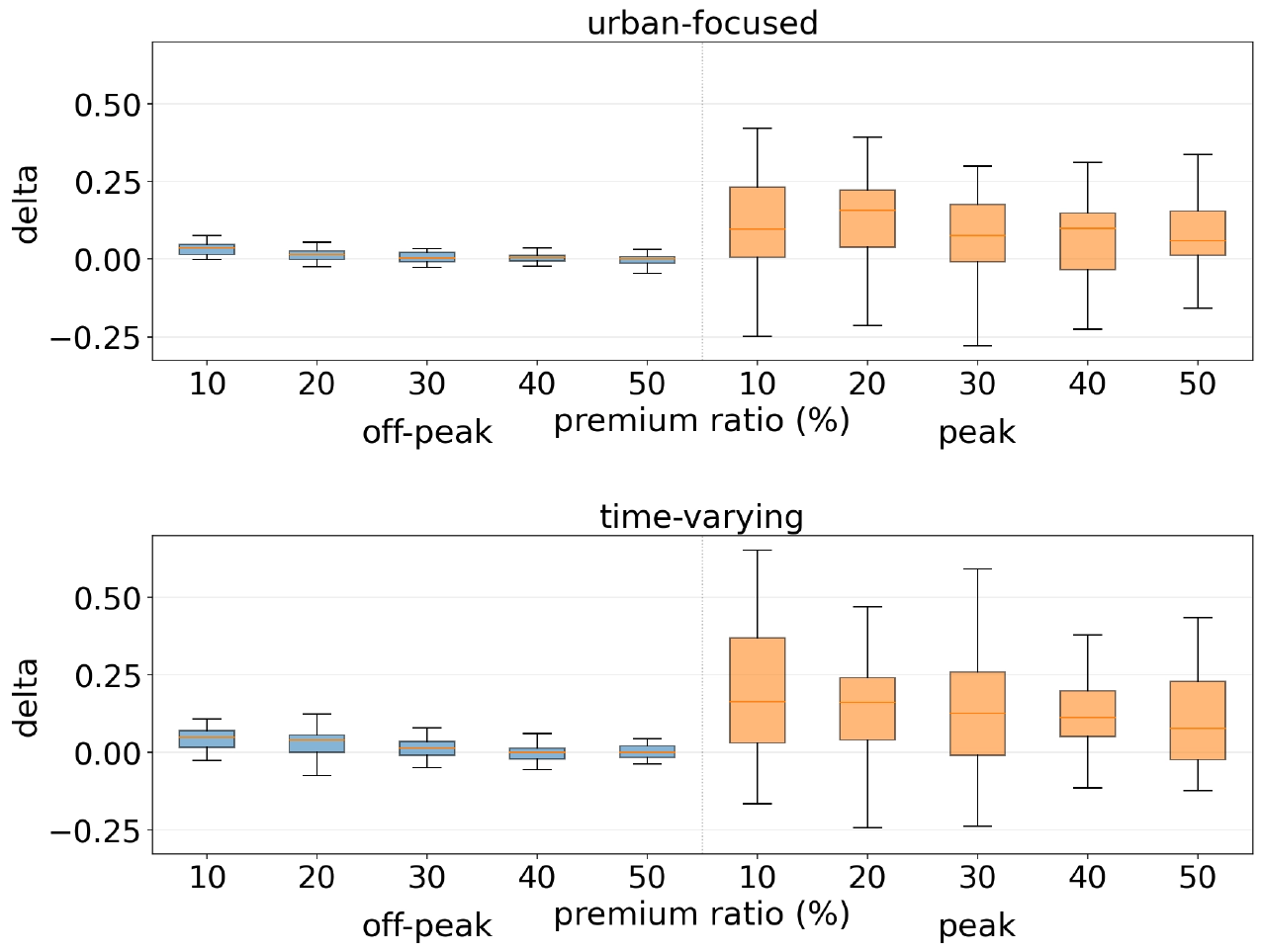}
\vspace{-0.5cm}
\caption{Distribution of $\delta$ for Shorter-Deadline in scenarios with peak periods.}
\label{fig:delta-shorter}
\end{figure}

Fig.~\ref{fig:delta-shorter} shows the 30-seed distribution of
$\delta$ for the scenarios with peak periods. The figure compares peak and off-peak periods and is used to examine whether \textit{Shorter-Deadline} can create a difference between non-premium and premium users.
The results show that \textit{Shorter-Deadline} induces a disparity in TPD between the two user groups during peak periods. This means that deadline shortening works as a priority rule when the fleet is under pressure. In contrast, $\delta$ becomes small during off-peak periods. A likely reason is that the number of vehicles is sufficient relative to demand, so many demands can be served almost one-to-one and ride-sharing occurs less often. 

\begin{figure}[tbp]
\centering
\includegraphics[width=0.75\textwidth]{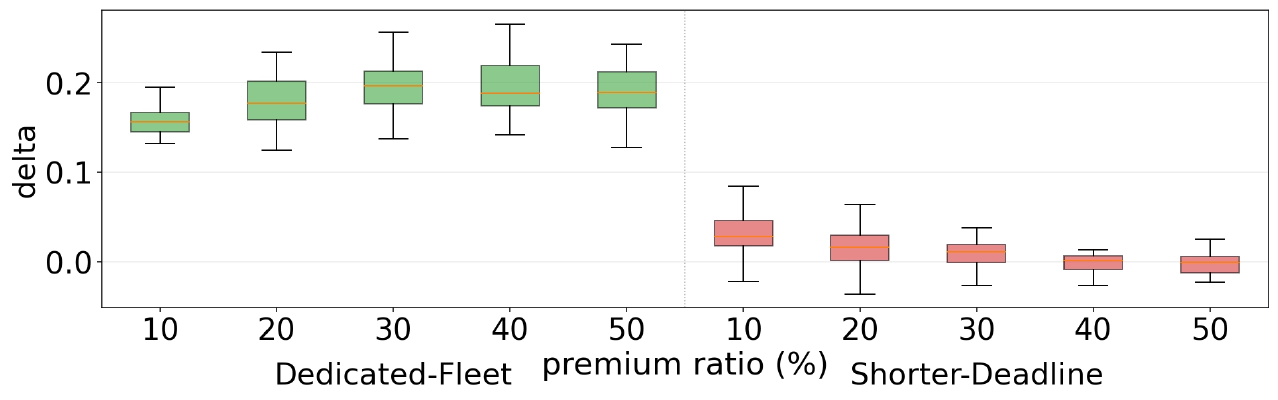}
\caption{Distribution of $\delta$ for \textit{Dedicated-Fleet} in the \textit{steady}.}
\label{fig:delta-exclusive}
\end{figure}

Fig.~\ref{fig:delta-exclusive} shows the 30-seed distribution of $\delta$ in the \textit{steady} without a peak period. The purpose is to examine whether dedicated premium vehicles can create a service gap in off-peak period.
The results indicate that introducing dedicated vehicles contributes to increasing $\delta$ under off-peak conditions. This is different from \textit{Shorter-Deadline}, whose effect becomes small when there is enough vehicle capacity. \textit{Dedicated-Fleet} creates a service gap by restricting non-premium users to shared vehicles while allowing premium users access to the entire fleet.

\subsection{Average Time per Distance for the Two User Groups}
\begin{figure}[tbp]
\centering
\includegraphics[width=\textwidth]{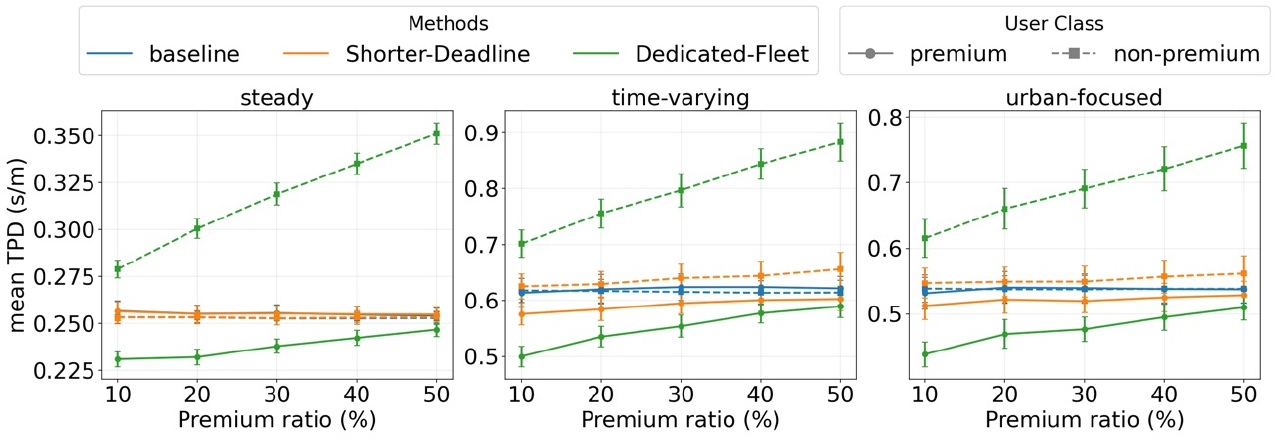}
\caption{Mean time per distance at $v_{\mathrm{prem}}=0.5$.}
\label{fig:mean-tpd}

\centering
\includegraphics[width=0.9\textwidth]{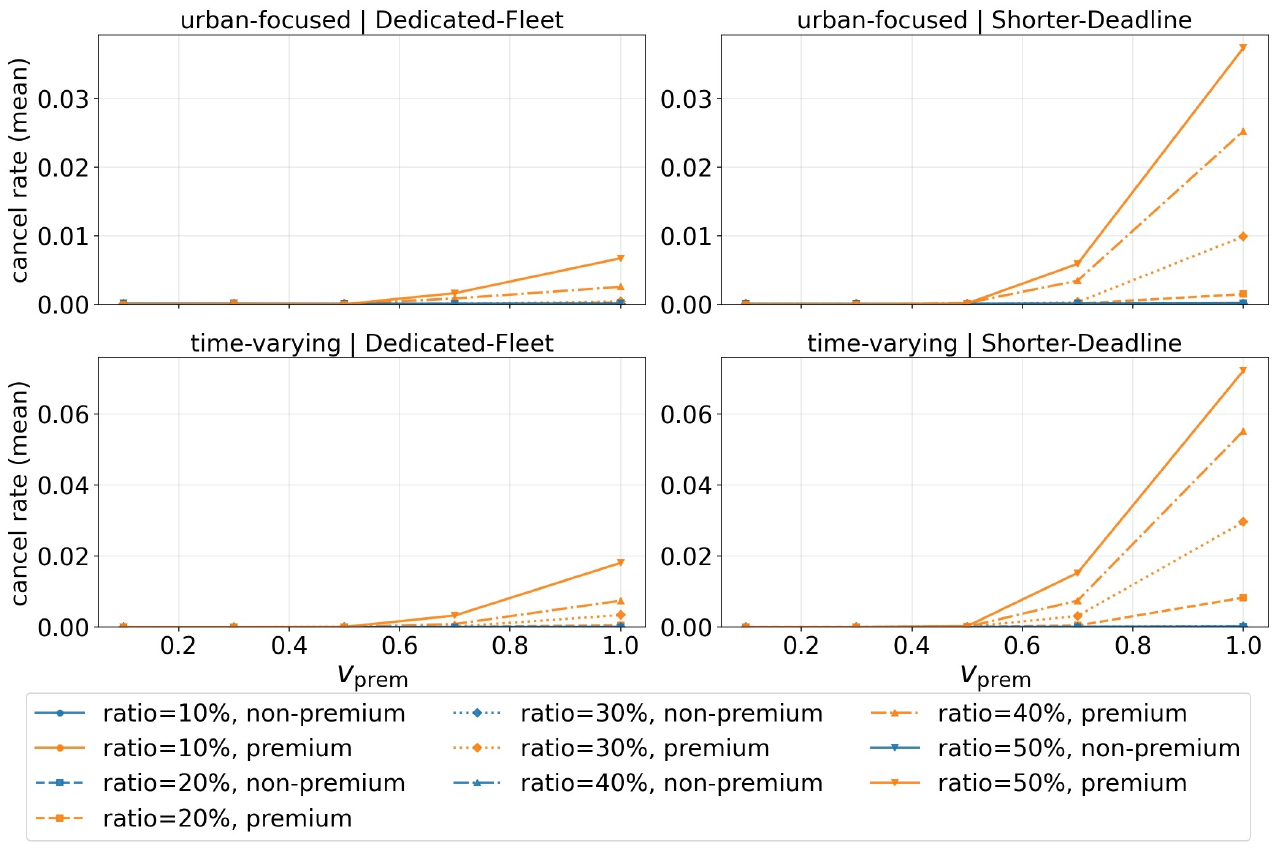}
\caption{Cancellation rate as a function of $v_{\mathrm{prem}}$. The cancellation rate remained zero under all conditions in \textit{Steady}. }
\label{fig:cancel}

\centering
\includegraphics[width=\textwidth]{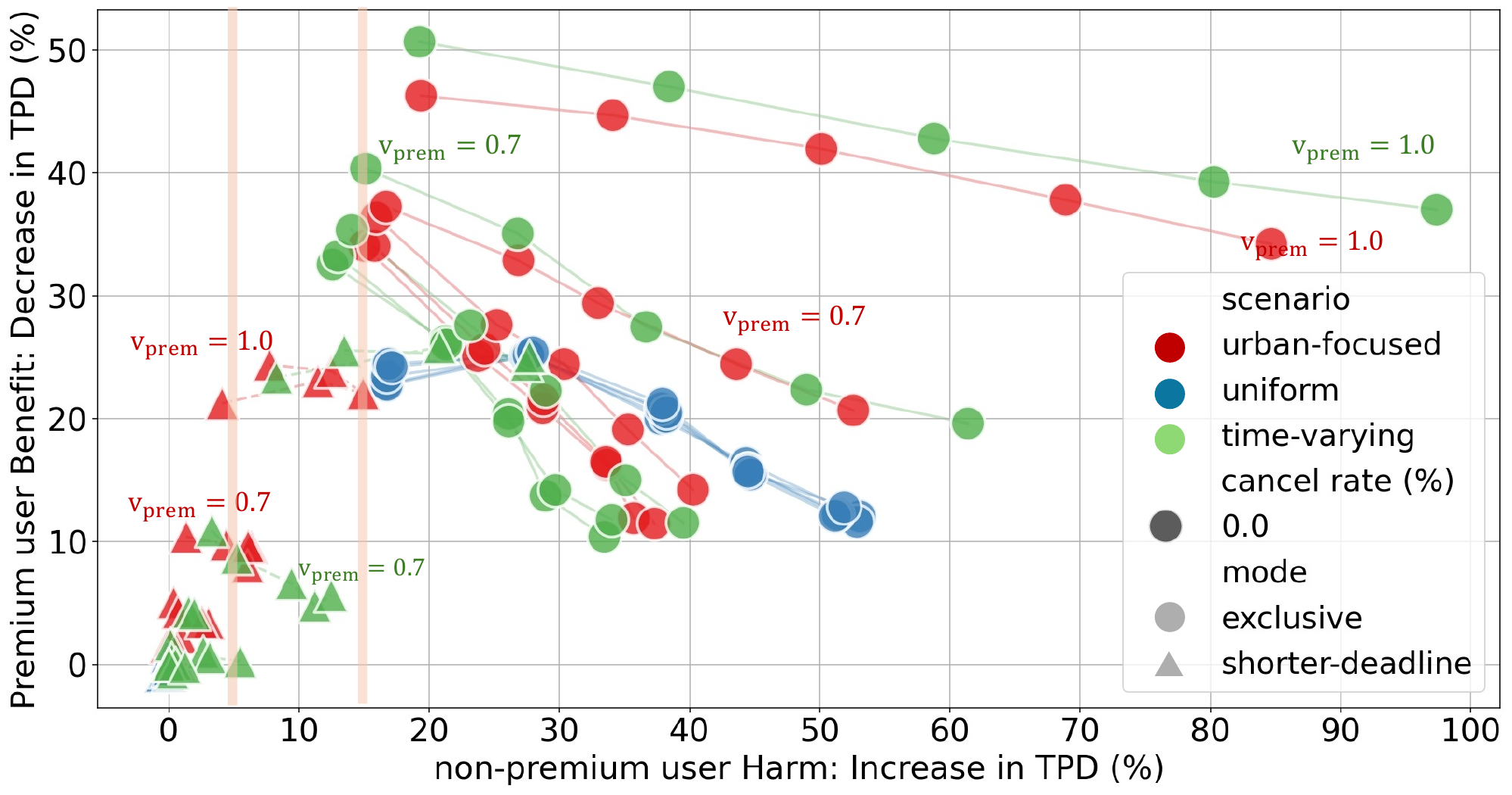}
\caption{Relationship between premium benefit and non-premium-user protection.Each polyline corresponds to one premium ratio, and points on a polyline correspond to different values of $v_{\mathrm{prem}}$.}
\label{fig:improve}
\end{figure}

Fig.~\ref{fig:mean-tpd} shows the mean TPD at $v_{\mathrm{prem}}=0.5$. 
The figure reports the 30-seed mean and 95\% confidence interval for premium and non-premium users under each scenario and method.
In the \textit{steady} scenario, differences among methods are limited. In the \textit{time-varying} and \textit{urban-focused} scenarios, higher premium ratios reduce the remaining benefit for premium users and increase the burden on non-premium users. This effect is stronger in \textit{Dedicated-Fleet}, because a larger premium ratio directly reallocates more vehicles away from non-premium users.



\subsection{Cancellation Rate and Improvement-Rate Space}

Fig.~\ref{fig:cancel} shows the cancellation rate for each scenario, dispatch method, and user group. The figure illustrates the relationship between $v_{\mathrm{prem}}$ and cancellations.
In \textit{Shorter-Deadline}, a larger $v_{\mathrm{prem}}$ makes the deadline stricter. As a result, more demands cannot be served before their shortened deadline, and the cancellation rate tends to increase. 
In contrast, \textit{Dedicated-Fleet} tends to keep the cancellation rate lower because dedicated vehicles absorb some strict premium demands.

Fig~\ref{fig:improve} shows the trade-off space between premium users' improvement rate and non-premium users' degradation rate from the baseline for each scenario, method, and $v_{\mathrm{prem}}$. 
Points are connected by $v_{\mathrm{prem}}$. Moving upward means that premium users receive a larger benefit, whereas moving rightward indicates greater degradation for non-premium users.
In \textit{Shorter-Deadline}, a U-turn pattern appears in the trade-off space under strict deadline settings such as $v_{\mathrm{prem}}=0.7$ and $1.0$. The triangle series sometimes shows 50\% cases with a smaller apparent non-premium burden than 30\% cases. Since TPD is computed only for completed trips, excessive cancellations can make the remaining trips appear more favorable. In \textit{Dedicated-Fleet}, dedicated vehicles absorb some strict premium demands, keeping cancellations lower and preventing this U-turn pattern.
The figure also shows a clear trade-off between premium user benefit and non-premium user impact. Among the tested points, limiting the increase in non-premium users' TPD to 5\% allows a maximum reduction of approximately 20\% in premium users' time per distance. When the acceptable deterioration for non-premium users is relaxed to 15\%, the maximum premium-user improvement among the tested points increases to approximately 30\%. Thus, rather than identifying a single universally optimal setting, the results suggest that the appropriate dispatch policy should be selected according to the operator's acceptable level of non-premium service degradation. 

\section{Discussion}

These results indicate that the proposed methods should be interpreted as different control policies whose suitability depends on operating conditions. \textit{Shorter-Deadline} preserves access to the full fleet for non-premium users and is useful when the fleet is under pressure. However, its effect becomes limited when capacity is sufficient. \textit{Dedicated-Fleet} can create a clearer service gap under off-peak conditions, but it can also reduce the effective capacity available to non-premium users under congestion.
This pattern appears across demand scenarios despite differences in spatial demand structure.

The results suggest that a fixed dedicated fleet is too rigid under congestion. When the system is congested, premium priority may be better implemented by deadline tightening through \textit{Shorter-Deadline}, while dedicated vehicles are more suitable under off-peak conditions. Thus, the results should be interpreted as guidance for choosing dispatch rules that create service differentiation while accounting for degradation among non-premium users.


\section{Conclusion}

We evaluated premium membership in on-demand taxi-sharing using two dispatch mechanisms: \textit{Shorter-Deadline}, which shortens the deadline for premium users in the sequential optimal insertion method, and \textit{Dedicated-Fleet}, which reserves part of the fleet for premium users while applying the same premium deadline rule. 

The results do not imply that premium membership is universally beneficial. Rather, they show that the suitable priority mechanism depends on operating conditions: \textit{Shorter-Deadline} is useful under congestion, whereas \textit{Dedicated-Fleet} is more suitable under off-peak conditions. 
Several open issues remain, including pricing, hybrid dispatch policies, non-random premium demand, additional demand scenarios, user-facing verification of priority, and validation under different map and fleet settings.

\section*{Acknowledgement}
This work was supported by JST-CREST Program Grant Number JPMJCR2564, Japan.

\bibliographystyle{splncs04}
\bibliography{refs}

\begin{thebibliography}{10}
\providecommand{\url}[1]{\texttt{#1}}
\providecommand{\urlprefix}{URL }
\providecommand{\doi}[1]{https://doi.org/#1}

\bibitem{AlonsoMora2017}
Alonso-Mora, J., Samaranayake, S., Wallar, A., Frazzoli, E., Rus, D.: On-demand high-capacity ride-sharing via dynamic trip-vehicle assignment. Proceedings of the National Academy of Sciences  \textbf{114}(3),  462--467 (2017). \doi{10.1073/pnas.1611675114}

\bibitem{Beirigo2022}
Beirigo, B.A., Negenborn, R.R., Alonso-Mora, J., Schulte, F.: A business class for autonomous mobility-on-demand: Modeling service quality contracts in dynamic ridesharing systems. Transportation Research Part C: Emerging Technologies  \textbf{136},  103520 (2022). \doi{10.1016/j.trc.2021.103520}

\bibitem{Chen2022TransitDesertsEurope}
Chen, Y., Jiao, J.: Are there transit deserts in europe? a study focusing on four european cases through publicly available data. Sustainability  \textbf{14}(20),  13182 (2022). \doi{10.3390/su142013182}

\bibitem{DeMunck2023}
De~Munck, T., Chevalier, P., Tancrez, J.S.: Managing priorities on on-demand service platforms with waiting time differentiation. International Journal of Production Economics  \textbf{266},  109053 (2023). \doi{10.1016/j.ijpe.2023.109053}

\bibitem{Fang2019PricesSubsidiesSharing}
Fang, Z., Huang, L., Wierman, A.: Prices and subsidies in the sharing economy. Performance Evaluation  \textbf{136},  102037 (2019). \doi{10.1016/j.peva.2019.102037}

\bibitem{HansenSener2023}
Hansen, T., Sener, I.: Strangers on this road we are on: A literature review of pooling in on-demand mobility services. Transportation Research Record: Journal of the Transportation Research Board  \textbf{2677},  036119812211238 (09 2022). \doi{10.1177/03611981221123801}

\bibitem{ITF2019AppBasedMobility}
{International Transport Forum}: Regulating app-based mobility services: Summary and conclusions. Tech. Rep.~175, OECD Publishing, Paris (2019). \doi{10.1787/94d27a3a-en}

\bibitem{Jiao2013TransitDeserts}
Jiao, J., Dillivan, M.: Transit deserts: The gap between demand and supply. Journal of Public Transportation  \textbf{16}(3),  23--39 (2013). \doi{10.5038/2375-0901.16.3.2}

\bibitem{Ma2024}
Ma, L., Tao, Z., Wei, Q.: On-demand ride service platform with differentiated services. PLOS ONE  \textbf{19},  1--27 (01 2024). \doi{10.1371/journal.pone.0296732}, \url{https://doi.org/10.1371/journal.pone.0296732}

\bibitem{noda2003evaluation}
Noda, I., Ohta, M., Shinoda, K., Kumada, Y., Nakashima, H.: Evaluation of usability of dial-a-ride systems by social simulation. In: International Workshop on Multi-Agent Systems and Agent-Based Simulation. pp. 167--181. Springer (2003)

\bibitem{deRuijter2021}
de~Ruijter, A., Cats, O., Alonso-Mora, J., Hoogendoorn, S.: Ride-pooling matching with a compensatory cost function: Implications for adoption, efficiency and level of service (2021), \url{https://arxiv.org/abs/2107.06710}

\bibitem{Santi2014}
Santi, P., Resta, G., Szell, M., Sobolevsky, S., Strogatz, S.H., Ratti, C.: Quantifying the benefits of vehicle pooling with shareability networks. Proceedings of the National Academy of Sciences  \textbf{111}(37),  13290--13294 (2014). \doi{10.1073/pnas.1403657111}

\bibitem{SozaParra2024}
Soza-Parra, J., Kucharski, R., Cats, O.: The shareability potential of ride-pooling under alternative spatial demand patterns. Transportmetrica A: Transport Science  \textbf{20}(2),  2140022 (2024). \doi{10.1080/23249935.2022.2140022}

\bibitem{WangZhang2022}
Wang, X., Zhang, R.: Carpool services for ride‐sharing platforms: Price and welfare implications. Naval Research Logistics (NRL)  \textbf{69} (10 2021). \doi{10.1002/nav.22030}

\end{thebibliography}

\end{document}